# Soft Photon Process and Redshifts of QSOs

Yi-Jia Zheng

National Astronomical Observatory, Chinese Academy of Sciences, Beijing, 100012 China

## Abstract

Quasars have reigned as the most luminous and distant objects in the Universe. These concepts are based on the hypothesis that the redshifts of quasars are obeying the Hubble law. This hypothesis has little serious competition today, but has met some difficulties when used to determine the distances of quasars.

However, the majority of the redshifts of quasars can be explained by the soft photon process proposed by Zheng (2013). As a consequence, quasars are not located at the distances that are calculated from observed redshifts according to the Hubble law. Therefore, distances of quasars can be reduced to within a reasonable range, and the difficulties that have been met by quasars will disappear.

It is now 50 years since quasars were identified. Since then quasars have reigned as the most luminous and distant objects in the Universe. These concepts are based on the hypothesis that the redshifts of quasars are obeying the Hubble law. This hypothesis has little serious competition today, but has met some difficulties when applied to derive the distances of QSOs.

However, the majority of the redshifts of quasars can be explained by the soft photon process proposed by Zheng (2013). For example, 3C 48 is the first identified quasar and has been investigated widely. Canalizo and Stockton (2000) have found that there are extended ionized gas and some concentrate clumps in the host galaxy of 3C 48. These findings were confirmed by Fu and Stockton (2009). In a brighter clump of extended emission region the electron number density $N_e$ has been found to be about 150 cm$^{-3}$, and the projected distance from the quasar $R$ has been found to be about 20 kpc. Even if the clump is not in hydrostatic pressure equilibrium with the surrounding medium, it is reasonable to assume that the ambient ionized gas of 3C 48 has a gas (?) number density $N$ about 10 cm$^{-3}$ and the extension along the line of sight $D$ is about of 10 kpc. Suppose the ambient gas is fully ionized because the gas mainly consists of hydrogen atoms. Then, the electron number density $N_e$ is approximately equal to the gas number density $N$. According to the soft photon process proposed by the author (Zheng 2013), the non-velocity redshift for 3C 48 can be estimated by $N_e D S_T$, where $S_T$ is the Compton scattering section. The value estimated is about of the order 1. Therefore, the majority of the redshift of 3C 48 may be caused by the soft photon process. Then, the cosmological redshift (in proportion to the distance of the quasar) of 3C 48 will be reduced to a much smaller amount compared with the observed redshift. Consequently, the distance of 3C 48 is not as large as derived from the Hubble law using the observed redshift.

If the distances of all quasars, like that of the 3C 48, can be reduced to within a reasonable range, then difficulties that have been met by quasars will be resolved easily.

1. Energy balances of quasars.

The discovery of radio variability and especially rapid intra-day variability in some quasars suggested correspondingly that quasars are of small linear dimensions. If the quasars are at cosmological distances derived from the observed redshifts according to the Hubble law, then the energy released by the quasars will be several orders larger than the ordinary galaxies. To balance the energy released by quasars some astronomers supposed that quasars involve at their core a super massive black hole producing energy by accretion of gas. But this hypothesis has not been confirmed satisfactorily. Taking the soft photon process into account, the distances of quasars can be reduced to within a reasonable range and energy balance problem of the quasars will disappear naturally.

2. The superluminal motions of components of quasars.

The structures of many quasars vary rapidly. These variations have been interpreted as superluminal expansions of two or more components. Taking 3C 345 as an example, an expansion velocity, $v/c$ =6.7, was derived (Wittels et al. 1976; Shafffer et al. 1977; Cohen et al. 1979). An important basis of this derivation is that the quasar 3C 345 is at the distance derived from the observed redshift according to the Hubble law. This is the same for the derivation of superluminal motions of components of all other quasars. If the distances of quasars are reduced to within a reasonable range, then the motions of the components of quasars drop correspondingly. The superluminal motions of components of quasars will be no longer exist.

3. Multiple redshifts of some individual quasar.

Arp et al. (1967) discovered that PKS 0273-23 has extraordinarily rich absorption-line spectrum. They fitted most of the absorption lines to a redshift of Z=2.201 as compared with an emission-line redshift of 2.223. Burbidge (1967) then observed PKS 0273-23 and identified a number of her measured absorption lines with a redshift of Z=1.95. Later, Greenstein and Schmidt (1967) obtained spectrograms having a higher resolution than those of previous authors and proposed that the both redshift systems, $Z_1$=2.2020 and $Z_2$=1.9555, were present. A systematic procedure has been used by Bahcall (1968) to identify twenty-eight of the lines with five absorption reshifts, namely, Z=2.2015, 1.6706, 1.6560, 1.5312, and 1.3642.

It is presumed that the absorption lines in PKS 0273-23 arise either in intergalactic space or in clouds which are near the object and moving at speeds of up to 0.3c relative to the source of the emission lines. At present it seems that most astronomers have advocated the former hypothesis that the absorption spectrum is formed in the intergalactic space and screened by absorbing objects, e.g., galaxies. The principal difficulty for the latter hypothesis is the huge redshift relative to the source of the emission.

Canalizo and Stockton (2000) have found that in the host galaxy of 3C 48 there are an extended distribution of the ionized gas and some concentrated clumps. This finding is in favor of the latter hypothesis that the absorption spectrum is formed in clouds which are near the object. The principal difficulty of the huge redshift relative to the source of the emission can be explained by the soft photon process proposed by Zheng (2013).

## Acknowledgment

I would like to thank Dr. Nailong Wu for the correction and suggestions his made to greatly improve the English of the manuscript.